\documentclass[12pt]{revtex4}
\usepackage{graphicx}
\usepackage{amssymb}
\begin{document}

\title{A laser-driven target of high-density nuclear polarized hydrogen gas}

\author{B. Clasie$^{1}$}
\author{C. Crawford$^{1}$}
\author{J. Seely$^{1}$}
\author{W. Xu$^{2}$}
\author{D. Dutta$^{2}$}
\author{H. Gao$^{1,2}$}
\affiliation{$^{1}$Laboratory for Nuclear Science, Massachusetts Institute of Technology, 
  Cambridge, MA 02139, USA}
\affiliation{$^{2}$Triangle Universities Nuclear Laboratory, Duke University, Durham, 
  NC 27708, USA}

%\date{\today}

\begin{abstract}
We report the best figure-of-merit achieved for an internal nuclear
 polarized hydrogen gas target and a Monte Carlo simulation of
 spin-exchange optical pumping.  The dimensions of the apparatus were
 optimized using the
 simulation and the experimental results were in good agreement with
 the simulation.
The best result achieved for this target was 50.5\% polarization 
with 58.2\% degree of dissociation of the sample beam exiting the
storage cell at a hydrogen flow rate of $1.1\times 10^{18}$ atoms/s.  

\end{abstract}
\maketitle

%\pacs{29.25.Pj, 34.50.Ez, 32.80.Bx}

The exploitation of polarization observables through the use of polarized
beams and polarized internal gas targets in storage rings is an increasingly 
valuable technique in nuclear and particle physics.
Nucleon properties, such as the spin structure 
functions and the electromagnetic form factors, have been measured in 
recent years
with polarization techniques utilizing polarized internal targets at DESY
(HERMES), NIKHEF and the MIT-Bates Laboratory.  
The spin-dependent asymmetry from the  
$\vec{p} + \vec{p} \rightarrow  p + p + \phi$ process has been 
suggested~\cite{titov00,titov98}
as a possible probe of the strangeness content of the nucleon. 
The near threshold $\vec{p} + \vec{p} \rightarrow Y + \Theta^{+}$ process
could be used to determine the parity of 
the $\Theta^{+}$ pentaquark 
state~\cite{thomas04,hanhart05,uzikov04},
if its existence is confirmed.

The Laser-Driven Target (LDT) is capable of producing nuclear polarized 
hydrogen and deuterium for storage rings. 
The LDT and the Atomic Beam Source (ABS) (another technique more 
commonly used) both use storage cells~\cite{steffens03} 
to increase the target thickness, 
compared to a free gas jet target.  However, the LDT 
offers a more compact design than the ABS, and can provide a 
higher Figure of Merit (FOM)~\footnote{The FOM is a measure of 
the performance of a polarized target, 
which determines the
statistical uncertainty of an asymmetry measurement for a given beam time.}
as reported in this work.
An LDT was first used in nuclear physics experiments~\cite{cadman01,miller98} 
in 1997 and 1998 at the Indiana University
Cyclotron Facility following earlier work on the laser-driven source and
target~\cite{holt89,poelker94,gao95,stenger96}.
A hydrogen LDT project was initiated 
at MIT in late 1990s with the goal of
implementing such a target in the South Hall Ring at the MIT-Bates 
Linear Accelerator Center for a precision measurement of the 
proton charge radius~\cite{gao1,gao2}.
In this paper we report the best FOM result obtained from this target, which
benefited from the development of a realistic Monte Carlo (MC)
simulation of the target.

A LDT is based on the technique of spin-exchange optical 
pumping. The valence electron of potassium 
is polarized through optical pumping in a magnetic field of $\sim$1~kG 
using circularly polarized laser light.   
%The potassium valence electron is excited from
%4$^{2}S_{1/2}$ to 4$^{2}P_{1/2}$ 
%followed by relaxation back to the 4$^{2}S_{1/2}$ state.  
%The helicity and wavelength 
%of the laser are chosen to excite one of the two magnetic substates, 
%resulting in depopulation of that substate and polarization 
%of the potassium species.
Spin exchange collisions then transfer the polarization from potassium 
to the Hydrogen (H) electron.  Finally, the hyperfine 
interaction during 
H-H collisions transfers the electron spin to the 
nucleus~\cite{happer72, wilbert_thesis}.  
If there are many
H-H collisions, the rate of transfer of spin to the nucleus equals the
reverse rate, and the system is in Spin Temperature Equilibrium 
(STE)~\cite{walker93}.
The time-constant for STE is approximately given by~\cite{walker93}:
\begin{equation}
\label{equation:ste}
\tau_{_{STE}} = \frac{1 +\left(B/B_c\right)^2}{n_{\mathrm{_{H}}}
  \sigma_{SE}^{\mathrm{HH}}v_{rel}^{\mathrm{HH}}},
\end{equation}
where $B_c$ is the critical magnetic field (507~G
for hydrogen), $n_\mathrm{H}$ is the density of atomic 
hydrogen (excluding molecular hydrogen), $\sigma_{SE}^{\mathrm{HH}}$ is
the thermally averaged H-H spin exchange cross section at  
the temperature of the spin-exchange cell and $v_{rel}^{\mathrm{HH}}$ is 
the average relative velocity 
between hydrogen atoms.
Laser-driven sources and targets are designed with the dwell time-constant in
the spin-exchange cell much greater
than the STE time-constant to guarantee that the system is in STE.
Moreover, STE has been verified in laser-driven sources and 
targets~\cite{stenger97,fedchak98,cadman_thesis}.  
Under STE conditions, 
the hydrogen nuclear and electron polarizations are equal~\cite{stenger97}.

\begin{figure*}[t]
  \centering
  \includegraphics[height=80mm]{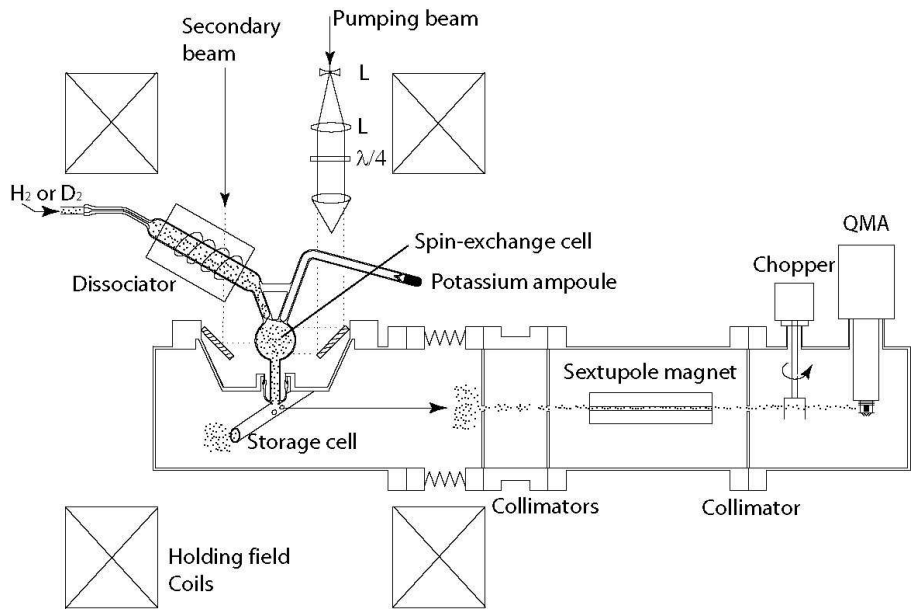}
  \caption{Laser Driven Target setup.  Note that, for clarity, the polarimeter 
    arm, storage cell, dissociator, and potassium ampoule are shown 
    rotated by 90$^{\circ}$ from their actual positions (in the 
    actual setup, the polarimeter arm and ampoule would come out of
    the page).  
An optional secondary beam
    can be used to measure the alkali density and polarization via a Faraday
    polarimeter.}  
  \label{fig:schematic-chris}
\end{figure*}

The main contributions to depolarization of the alkali and hydrogen 
in our apparatus come from the flow of atoms into and out of the 
LDS and depolarization during wall
collisions~\cite{stenger95}.  
Atoms may also recombine at a surface producing
molecules with predominantly zero net nuclear spin.  The 
recombination is characterized by the degree of dissociation,
f$_\alpha$, which is the fraction of the hydrogen flux in the sample
beam exiting the storage cell that is in atomic form.
Drifilm coatings are employed to limit the recombination and 
depolarization effects from wall 
collisions~\cite{fedchak97}.  
The depolarization from radiation
trapping~\cite{tupa87-1, tupa87-2} 
can be limited by optical pumping in a large magnetic field in the kG
range~\cite{walker93}; however, the rate of transfer of spin 
to the nucleus by the 
hyperfine interaction is reduced.  
Therefore, the magnitude of the magnetic
field in the spin-exchange cell must be optimized for these
two competing effects~\cite{walker93,anderson95,stenger95}.

Figure~\ref{fig:schematic-chris} is a schematic view of the MIT LDT.
Hydrogen gas flows successively into different sections of a 
piece of pyrex glassware (which consists of a dissociator tube, 
a spin-exchange cell and a transport tube) and an aluminum 
storage cell.  The molecular 
gas is dissociated into atoms by an RF discharge in the dissociator tube.  
In the spin-exchange cell the hydrogen gas (now a mixture of atoms and 
molecules) is mixed with the potassium vapor 
produced in a side-arm by heating a potassium ampoule.
The results from two spin-exchange cells, ``Original'' and
``Large-1'', are reported herein.  
To minimize the number of wall collisions, the Original spin-exchange
cell design was spherical with an inner
diameter of 4.8~cm.  Large-1 was a cylindrical cell optimized by the
MC simulation described below.
The entire volume contained by the spin-exchange cell, transport
tube and storage cell must be heated to 200--250$^\circ$C to 
prevent the alkali-metal vapor
from condensing on the walls, which would degrade the drifilm coating.  
The potassium number density
is typically 0.3\% compared to H.  
The standard storage cell is 
an open ended aluminum cylinder coated with drifilm.  The cell is 40~cm 
in length, and 1.25~cm in diameter with  two sampling holes 
allowing the target gas to 
be monitored by an atomic polarimeter.
One hole is centered and the other one is 15~mm downstream. Both 
are positioned at right angles to the entrance hole of the storage cell, 
which ensures that the atoms monitored by the 
atomic polarimeter undergo wall collisions in the storage cell before 
escaping the cell.
A MC simulation determined that atoms that exit the center 
(off-center) sampling hole
experience, on average, 1370 (1370) wall collisions of which 135 (155) 
wall collisions are in the storage cell. 

The laser used 
is a Titanium-Sapphire laser 
(Ti:Sapph) pumped with a 20~W Argon ion 
laser.  The laser beam passes 
through an Electro-Optic Modulator (EOM, not shown), an expanding
lens, and a quarter-wave plate before arriving at the spin-exchange cell via 
a periscope with two polarization-preserving mirrors.  
The EOM broadens the relatively narrow linewidth of
the Ti:Sapph laser to provide a better match to the 
potassium Doppler absorption 
profile with a FWHM of 1.0~GHz. 
In addition,
two sampling beams are split off from the pump beam for monitoring the laser
spectrum and wavelength.  

Gas exiting the sampling hole of the storage cell is collimated through a
series of apertures which also serve as conductance limiters between
sub-chambers of the polarimeter.  A permanent sextupole magnet focuses
one electronic spin state of the atomic beam and defocuses the other.  The
optimal focal length was determined by an atomic beam simulation.
The beam is then sampled by a Quadrupole Mass Analyzer (QMA)  
which alternately measures both the atomic
and molecular intensities.  The QMA is shielded from the holding field by two
layers of $\mu$-metal.  The small signal at $\sim$1~m from the storage 
cell is enhanced using a chopper along with a lock-in amplifier.  
The background pressure is reduced to $10^{-9}$~Torr by differentially
pumping the two sub-chambers with ion pumps and also a NEG pump in the second
(QMA) chamber.  The background can be measured by blocking the beam with a
shutter or moving/rotating the polarimeter away from the sampling hole.  

The degree of dissociation of the sample beam exiting the storage cell
is given by the change in the 
molecular signal (after subtracting the background) 
when the RF discharge is turned on and off. 
The electron polarization of the atomic hydrogen species, $P_e$,
is given by the change in the atomic signal when
the laser is turned on and off 
by opening or closing a laser
shutter (after subtracting the background).
This measurement also indicates the hydrogen 
nuclear polarization, as the system is designed to be 
in STE.  The mean dwell time for atomic hydrogen in the 
Original spin-exchange cell and transport tube 
has been calculated by a MC simulation, 
to be 8.8~ms.  For a field of 100~mT
and an atomic hydrogen density of $1.0\times 10^{14}~\mathrm{atoms/cm^3}$ 
the time-constant for STE given by Equation \ref{equation:ste} is 0.052~ms.  
The mean dwell time is therefore larger than the STE time
constant by a factor of approximately 170.  
The Erlangen hydrogen LDS was verified 
to be in STE by directly measuring
the nuclear polarization \cite{stenger97} in conditions where the mean 
dwell time was larger than the STE 
time-constant by a factor of 300, and the system was expected to remain in STE 
at half that ratio.

\begin{table}[t]
  \begin{center}
     \begin{ruledtabular}
       \begin{tabular}{|l|c|c|c|c|c|c|}
         & \multicolumn{2}{c}{HERMES}\vline
         & \multicolumn{2}{c}{IUCF}  \vline
         & \multicolumn{2}{c}{MIT LDT} \vline\\
         & \multicolumn{2}{c}{(ABS)} \vline
         & \multicolumn{2}{c}{(LDT)} \vline
         & Original & Large-1 \\
         \hline
         Gas                             &H     &D     &H          &D           &H      &H      \\
         F                               &6.57 &5.15   &100        &72          &110    &110    \\
         t                               &11    &[10.5]&50         &50          &150    &150    \\
         f$_\alpha$                      &      &      &$\sim$0.48 &$\sim$0.48  &0.56   &0.58   \\
         $P_e$                           &      &      &$\sim$0.45 &$\sim$0.45  &0.37   &0.50   \\
         $\langle p_z\rangle$            &0.78  &0.85  &0.145      &0.102       &[0.175]&[0.247]\\
         F$\times\langle p_z\rangle^2$   &4.0   &3.8   &2.1        &0.75        &3.4    &6.7    \\
         t$\times\langle p_z\rangle^2$   &6.7   &7.6   &1.1        &0.52        &4.6    &9.2    \\ 
       \end{tabular}
     \end{ruledtabular}
  \end{center}
  \caption{FOM results from the HERMES ABS~\cite{airapetian05,hermes_tech_h, hermes_tech_d}, IUCF
  LDT~\cite{cadman_thesis}, and the MIT LDT. The units are as follows; 
  the flow, F ($10^{16}$ atoms/s); 
  the thickness, t ($10^{13}$ atoms/cm$^2$); 
  the FOM, F$\times\langle  p_z\rangle^2$ ($10^{16}$ atoms/s); 
  and, the FOM, t$\times\langle  p_z\rangle^2$ ($10^{13}$ atoms/cm$^2$).  
%  For the MIT LDT, Large-1 is the improved spin-exchange cell design.
  All LDT results for f$_\alpha$ are under operating
  conditions, with the potassium ampoule heated. 
%  Numbers inside square
%  brackets are described further in the text.
  }
  \label{tab:results}
\end{table}

\begin{figure}[!h]
  \centering
  \includegraphics[width=80mm]{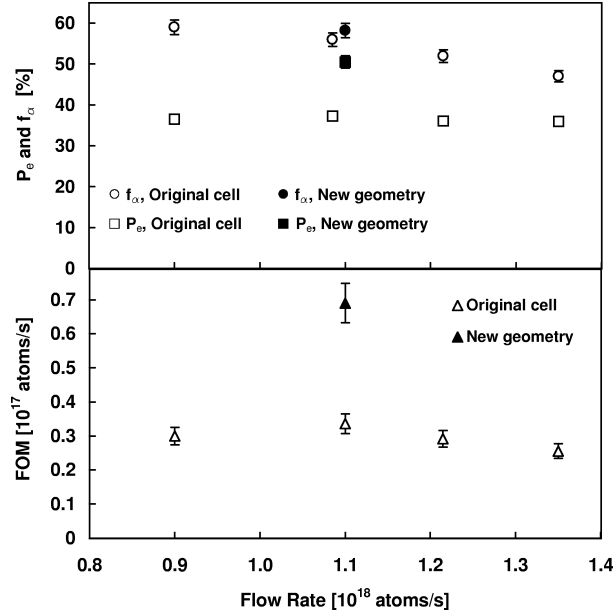}
  \caption{Results achieved by the MIT LDT. 
    The FOM is given as
    flow$\times\langle p_z\rangle^2$
    where $\langle p_z\rangle$ is the density averaged H nuclear vector
    polarization assuming the system is in STE.} 
 
  \label{fig:results}
\end{figure}

Results from the IUCF and MIT LDTs are summarized in Table~\ref{tab:results} 
along with
results from the HERMES ABS.  The FOM is given as
flow$\times\langle p_z\rangle^2$ and thickness$\times\langle p_z\rangle^2$,
where $\langle p_z\rangle$ is the density averaged nuclear vector polarization.
For the MIT LDT, 
\begin{equation} \label{equation:pz_equation}
\langle p_z\rangle = {\frac{{\rm f}_\alpha P_e}{{\rm f}_\alpha +
    \sqrt{2}(1-{\rm f}_\alpha)}}.
\end{equation}
For the IUCF target, $\langle p_z\rangle$,
was determined from a scattering
experiment~\cite{cadman_thesis,cadman01}. 
The results for the flow and target thickness of the HERMES ABS are
based on Refs.~\cite{airapetian05,hermes_tech_h,hermes_tech_d}, and
are the best published ABS results that use a storage cell. 

The best results achieved by the MIT LDT in the 
Original configuration 
are shown in Fig.~\ref{fig:results} 
together with the overall errors which are dominated by the 
systematic errors.  
 The combined systematic uncertainty in the FOM is
estimated to be 8.4\%, which is dominated by the 
non-linearity in the QMA
response (less than 3\%), hydrogen flow control (4\%), and 3\% in 
the density averaged target polarization.
The measurements were 
repeated several times for this cell geometry under various conditions,
including recoating the surface, with reproducible results.

A detailed MC simulation of optical pumping 
and spin-exchange collisions
for our target was developed and used to extract the  
recombination and depolarization coefficients, and 
to provide a new cell design to
improve the target performance.  
The simulation techniques developed for the LDT are also applicable to
the design of the ABS, particularly at future facilities 
where constraints may cause significant recombination and/or
spin-exchange.  The recombination coefficient,
$\gamma_r(n_{\mathrm{H}})$, is the probability for a hydrogen atom to
recombine at a wall collision and is a function of $n_{\mathrm{H}}$
near the surface.  As $n_{\mathrm{H}}$ varies 
throughout the simulated volume, $\gamma_r$ changes with position
on the surfaces.

In the simulation, a hydrogen atom
moves ballistically between wall 
collisions in the spin-exchange cell, transport tube and storage cell.
A new velocity, both magnitude and direction, is randomly generated
after each wall collision, according to a Maxwellian and a $\cos
\Theta$ distribution, respectively, where $\Theta$ is the polar angle
measured with respect to the normal to the surface. At high
temperatures, which are experienced in an LDT, the recombination
coefficient, $\gamma_r$, is given by~\cite{hauke_thesis}
\begin{equation} \label{equation:recomb}
  \gamma_r(n_{\mathrm{H}}) = C_{\mathrm{H}}n_{\mathrm{H}},
\end{equation}
where $C_{\mathrm{H}}$ is a constant.
After the hydrogen atom exited either through a sampling hole or the
ends of the storage cell, another hydrogen atom was generated at the
top of the spin-exchange cell. 

The MC was used to determine $n_{\mathrm{H}}$ throughout the
apparatus.  As $n_{\mathrm{H}}$ depends on the average probability
that atoms have not recombined at a given point, and this probability
depends on $n_{\mathrm{H}}$ through Equation~\ref{equation:recomb},
the simulation was iterated and $n_{\mathrm{H}}$ 
recalculated after each iteration until the degree of dissociation of 
atoms exiting the storage cell sampling hole and $n_{\mathrm{H}}$
converged. 

Although the hydrogen atoms were transported separately, H-H and H-K
spin-exchange collisions were treated by allowing the hydrogen atoms to
interact with the average hydrogen electron and nuclear polarization
and potassium electron polarization.
The apparatus was divided into a 3-dimensional $2\times2\times2$~mm$^3$
grid.  Initial values of the average H and K polarizations were assigned at
every point on the grid.  After a hydrogen atom exited the apparatus in
the simulation, the average polarizations were updated, and the
simulation was iterated until convergence.  A hydrogen atom can be 
depolarized during a wall collision with the probability given by the
%depolarization coefficient, $\gamma_p = 0.00355$.  
depolarization coefficient, $\gamma_p$.  
The MC results were fit to the experimental results for the
Original configuration, shown in Table~\ref{tab:results}, by varying
$C_{\mathrm{H}}$ and $\gamma_p$, which were determined to be
$3.33\times10^{-18}~{\rm cm}^3$ and $0.00355$ respectively.
Further discussion of the MC simulation will be
reported in a forthcoming paper.

A cylindrical spin-exchange cell with a much larger diameter 
was constructed based on the MC studies and the practical constraints
of our target chamber.  The calculated mean dwell time
divided by the STE time constant was 280. This design, labeled Large-1 
in Table~\ref{tab:results}, has a spin-exchange cell volume 6.8 
times larger than that of the Original design.
%% add some description of the new cell geometry
The best result obtained for this cell using the EOM
was $P_e=50.5\%$ and f$_\alpha= 58.2\%$
at a flow rate of $1.1\times 10^{18}$~atoms/s.
These results are in good agreement with the MC
simulation, which predicted $P_e=57\%$ and f$_\alpha= 51\%$ at the
same flow rate.

While in the Original configuration, drifilm coatings were found to
last in excess of 100 hours under operating conditions. 
The polarization result for the Large-1 cell
was stable at 50\% polarization for about 12 hours but with rather rapid 
deterioration of the dissociation fraction. This observation
may have been due to uneven heating of the spin-exchange cell
and the transport tube. For the Large-1 geometry, there was 
only a 1~cm gap for the
hot air to circulate around the glass due to the constraint of the existing
target chamber. One can overcome this constraint with the design of 
a new target chamber.  

One may argue that it is probably not completely justified to compare the 
performance of the HERMES ABS target and the IUCF LDT with the FOM of 
the LDT obtained in our polarized target lab due to the difference in
the storage cell conductances. A detailed study 
shows that minimal modifications are needed for the installation of this target
in the MIT-Bates storage ring.
A more realistic comparison which takes into account
correction factors due to the target geometry, temperature
and molecules still shows that our target with the Large-1 cell geometry has 
$(33 \pm 11)\%$ higher figure of merit than that of the HERMES hydrogen target.
These results represent an even larger improvement compared to the
previous best FOM from an LDT, which was obtained at IUCF.  
A similar comparison that does not bias toward the MIT LDT 
due to the storage cell conductance is $(210 \pm 30) \%$ higher 
than the IUCF result.
These comparisons
will be explained in a forthcoming paper. 

We thank Tom Wise and Willy Haeberli for the construction of the storage
cells; Michael Grossman and George Sechen for their technical
support; Ernest Ihloff, Manouchehr Farkhondeh, William Nispel and Defa Wang 
for their help with the vacuum chamber, the laser system,
fabrication of the spin-exchange cell oven, and the RF system;
Tom Hession for the fabrication of the spin-exchange cells;
and T. Black for his help in the early stage of this project.
We appreciate the useful discussions with Hauke Kolster.  
We thank J. Stewart and P. Lenisa for the information on the HERMES ABS target.
This work is 
supported in part by the U.S. Department of Energy under contract number 
DE-FC02-94ER40818.  H.G. acknowledges the support of an Outstanding Junior 
Faculty Investigator Award from the DOE.

\end{document}